\documentclass[manuscript, acmsmall]{acmart}

\usepackage{listings}
\usepackage{xcolor}
\usepackage{cleveref}
\usepackage{subcaption}
\usepackage{braket}

\crefname{figure}{Figure}{Figures}
\newcommand{\subfigref}[1]{Figure~\ref{#1}}

\renewcommand\footnotetextcopyrightpermission[1]{} 
\begin{document}

\title{Qrisp Implementation and Resource Analysis of a T-Count-Optimised Non-Restoring Quantum Square-Root Circuit}

\author{Heorhi Kupryianau}
\email{heorhi.kupryianau@tele.agh.edu.pl}

\author{Marcin Niemiec}
\email{marcin.niemiec@agh.edu.pl}
\affiliation{%
  \institution{AGH University of Krakow}
  \city{Krakow}
  \country{Poland}
  \address{Mickiewicza 30, 30-059}
}

\begin{abstract}
\noindent Efficient quantum arithmetic operations are essential building blocks for complex quantum algorithms, yet few theoretical designs have been implemented in practical quantum programming frameworks. This paper presents the first complete implementation of the T-count optimized non-restoring quantum square root algorithm using the Qrisp quantum programming framework. The algorithm, originally proposed by Thapliyal et al., offers better resource efficiency compared to alternative methods, achieving reduced T-count and qubit requirements while avoiding garbage output. Our implementation validates the theoretical resource estimates, confirming a T-count of $14n-14$ and T-depth of $5n+3$ for $n$-bit inputs. The modular design approach enabled by Qrisp allows construction from reusable components including reversible adders, subtractors, and conditional logic blocks built from fundamental quantum gates. The three-stage algorithm—comprising initial subtraction, iterative conditional addition/subtraction, and remainder restoration is successfully translated from algorithmic description to executable quantum code. Experimental validation across multiple test cases confirms correctness, with the circuit producing accurate integer square roots and remainders. This work demonstrates the practical realizability of resource-optimized quantum arithmetic algorithms and establishes a foundation for implementing different arithmetic operations in modern quantum programming frameworks.
\end{abstract}


\maketitle

\section{Introduction}
Quantum computing has emerged as a paradigm capable of solving certain problems much faster than classical computers. Quantum algorithms use superposition and entanglement to achieve these speedups. For example, Shor's algorithm \cite{Shor1997} for integer factorization runs in polynomial time and capable of breaking popular cryptographic algorithms like RSA. Grover's search on the other hand can find a target item in an unsorted database in $O(\sqrt{N})$ steps instead of $O(N)$ that can speed up brute force attacks \cite{Grover1997}. More broadly, quantum algorithms have been developed for applications ranging from cryptography to linear algebra and the simulation of physical systems \cite{Montanaro2016} \cite{Harrow2009} \cite{pawlitko2025}. These advances highlight the transformative potential of quantum computing across diverse fields of science and engineering.

Among the computational tasks that quantum computers will tackle, implementing fundamental arithmetic operations is crucial for enabling larger algorithms. One such operation is the square root, which is common in scientific and engineering computations. Square root calculations appear in contexts ranging from numerical methods and signal processing to certain cryptographic protocols and higher-level algorithms \cite{Wang_2020} \cite{Cheung2008} \cite{Bhaskar2016} \cite{Beauregard2003}. Accordingly, efficient quantum circuits for functions like the square root are needed to integrate quantum computing into these applications.

Several quantum circuit designs for the square root operation have been proposed in the literature, each with varying trade-offs between gate count, qubit usage, and garbage output. One particularly efficient approach is based on the non-restoring square root algorithm \cite{Thapliyal:2018}. This algorithm has been shown to yield a quantum circuit with reduced $T$-count and qubit requirements compared to other square root methods, such as those based on Newton iteration \cite{Bhaskar2016}. The design minimizes ancilla usage and avoids garbage output by construction, and its resource efficiency has been analyzed in detail in terms of both gate complexity and fault-tolerant considerations.

In this work, we implement the non-restoring square root algorithm proposed in \cite{Thapliyal:2018} using the Qrisp quantum programming framework \cite{seidel2024qrisp}. Qrisp enables high-level construction of reversible circuits and offers built-in support for common quantum operations, which are used in the implementation of arithmetic components. The modularity and flexibility of Qrisp allow for a direct mapping of the algorithm's structure into a circuit composed of reusable subcomponents, including reversible addition, subtraction, and conditional logic blocks.

The implementation shows that the optimized algorithm from \cite{Thapliyal:2018} can run in a modern quantum framework. We built and tested the circuit in Qrisp, ran simulations, and measured its resource use under realistic settings. This work proves that low-overhead arithmetic designs can be turned into working quantum circuits using tools such as Qrisp.

In the following, we give a roadmap for the remainder of this paper. Section \nameref{basics} reviews the quantum gates used in the algorithm and the theoretical background. \nameref{quantum_square_root} introduces the quantum square root algorithm that is being implemented. \nameref{implementation} section presents the implementation of the key parts of the algorithm in Qrisp. \nameref{validation} reports validation experiments and the performance results. Finally, \nameref{conclusion} concludes with a summary of our contributions and prospects for future work.

\section{Basics}
\label{basics}

This section briefly introduces selected types of quantum gates and circuits which are needed to describe and build quantum square root algorithm. 

\subsection{T-Depth and T-Count}

Let a quantum circuit be expressed over the Clifford+T gate set. Two common cost metrics are:

\begin{itemize}
  \item \textbf{T-Count:} Total number of $T$ gates in the circuit.
  \item \textbf{T-Depth:} Minimum number of sequential layers of $T$ gates, where gates in the same layer act on disjoint qubits and may be executed in parallel.
\end{itemize}

These metrics are used because $T$ gates are expensive to implement in fault-tolerant quantum computation. Optimizing for low T-Count reduces the total overhead, while optimizing T-Depth minimizes the circuit runtime. Together, they help assess the practicality of quantum algorithms on real hardware.

\subsection{The NOT Gate}
The NOT gate is a single qubit gate represented as shown in \subfigref{fig:not_gate}. Since it does not contain $T$ gates, its $T$-count and $T$-cost is $0$.

\subsection{The CNOT Gate}
\begin{figure}[h]
    \Description[NOT and CNOT gates]{Quantum circuit diagrams showing the NOT gate which inverts a single qubit and the CNOT (Controlled-NOT) gate which performs conditional bit flip on a target qubit based on a control qubit.}
    \begin{subfigure}[t]{0.18\textwidth}
        \includegraphics{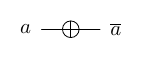}
        \subcaption{NOT gate.}
        \label{fig:not_gate}
    \end{subfigure}
    \hspace{1em}
    \begin{subfigure}[t]{0.22\textwidth}
        \includegraphics{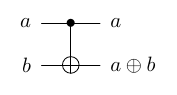}
        \subcaption{CNOT gate.}
        \label{fig:cnot_gate}
    \end{subfigure}
    \caption{NOT and CNOT gates.}
    \label{fig:not_and_cnot_gates}
\end{figure}

The Controlled-NOT (CNOT) gate is a 2-qubit reversible gate having the mapping for input qubits $a$ and $b$ to output qubits $a$ and $a \oplus b$ respectively. Quantum represented of the CNOT gate is shown in \subfigref{fig:cnot_gate}. $T$-count and $T$-cost of the CNOT gate is $0$.

\subsection{The SWAP Gate}

\begin{figure}[h]
    \Description[SWAP gate and its decomposition]{Quantum circuit representation of the SWAP gate which exchanges states between two qubits, along with its equivalent implementation using three CNOT gates.}
    \begin{subfigure}[t]{0.18\textwidth}
        \includegraphics{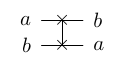}
        \subcaption{SWAP gate.}
        \label{fig:swap_gate}
    \end{subfigure}
    \hspace{1em}
    \begin{subfigure}[t]{0.3\textwidth}
        \includegraphics{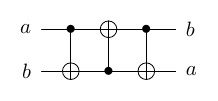}
        \subcaption{Decomposition of the SWAP gate.}
        \label{fig:swap_gate_decomposition}
    \end{subfigure}
    \caption{The SWAP gate.}
    \label{fig:swap_gate_group}
\end{figure}

The SWAP gate is a gate that swaps the states of two qubits. A quantum representation of the gate in quantum circuit is shown in \subfigref{fig:swap_gate}. It could be decomposed into 3 CNOT gates as shown in \subfigref{fig:swap_gate_decomposition}. $T$-count and $T$-cost of the CNOT gate is $0$.

\subsection{The Hadamard Gate}

\begin{figure}[h]
    \Description[Hadamard gate transformations]{Quantum circuit showing the Hadamard gate which creates superposition states, transforming basis states into equal superpositions with different phases.}
    \includegraphics{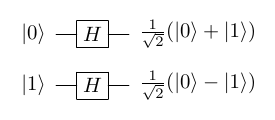}
    \caption{The Hadamard gate.}
    \label{fig:hadamard_gate}
\end{figure}

The Hadamard gate is a single qubit gate that maps the state $\ket{0}$ to the superposition state $\frac{1}{\sqrt{2}}(\ket{0} + \ket{1})$ and the state $\ket{1}$ to the superposition state $\frac{1}{\sqrt{2}}(\ket{0} - \ket{1})$. A quantum representation of the Hadamard gate is shown in \ref{fig:hadamard_gate}. The Hadamard gate is a Clifford gate; therefore, the $T$ count and $T$ depth of the Hadamard gate are $0$.

\subsection{The T and T\texorpdfstring{$^\dagger$}{†} Gates}

The T and T\texorpdfstring{$^\dagger$}{†} gates are single qubit gates that are used to control the phase of a qubit. The T gate is a gate that adds a phase of $\frac{\pi}{4}$ to the state and the T\texorpdfstring{$^\dagger$}{†} gate is a gate that removes a phase of $\frac{\pi}{4}$ from the state.

\subsection{The Toffoli Gate}

\begin{figure}[h]
    \Description[Toffoli gate circuit]{Quantum circuit diagram of the Toffoli (CCNOT) gate showing both its standard representation and its decomposition into Hadamard, CNOT, T and T-dagger gates.}
    \begin{subfigure}[t]{0.27\textwidth}
        \includegraphics{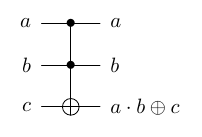}
        \caption{Toffoli gate graphical representation.}
        \label{fig:toffoli_gate}
    \end{subfigure}

    \begin{subfigure}[t]{0.65\textwidth}
        \includegraphics{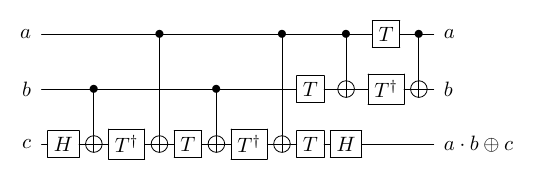}
        \caption{Toffoli gate decomposition.}
        \label{fig:toffoli_gate_decomposed}
    \end{subfigure}

    \caption{The Toffoli gate.}
    \label{fig:toffoli_gate_group}
\end{figure}

The Toffoli gate is a 3-qubit reversible gate having the mapping for three input qubits $(a, b, c)$ to three output qubits $(a, b, a.b \oplus c)$ as shown in \subfigref{fig:toffoli_gate}. 

One of the Toffoli gate realization is presented in \subfigref{fig:toffoli_gate_decomposed} and consists of two Hadamard gates, six CNOT gates, four $T$ gates and four $T^\dagger$ gates. Therefore, the $T$-count of the Toffoli gate is $7$ and the $T$-depth is $3$.

\subsection{The Peres Gate}

\begin{figure}[h]
    \Description[Peres gate circuit]{Quantum circuit showing the Peres gate which combines Toffoli and CNOT operations, presented both in its compact form and decomposed implementation.}
    \begin{subfigure}[h]{0.25\textwidth}
        \includegraphics{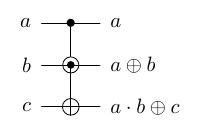}
        \subcaption{Peres gate graphical representation.}
        \label{fig:peres_gate}
    \end{subfigure}
    \hspace{1em}
    \begin{subfigure}[h]{0.3\textwidth}
        \includegraphics{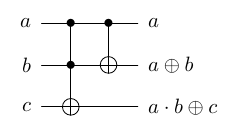}
        \subcaption{Peres gate decomposition.}
        \label{fig:peres_gate_decomposed}
    \end{subfigure}
    \caption{The Peres gate.}
    \label{fig:peres_gate_group}
\end{figure}

The Peres gate is a 3-qubit reversible gate having the mapping for three input qubits $(a, b, c)$ to output qubits $(a, a \oplus b, a \cdot b \oplus c)$ as shown in \subfigref{fig:peres_gate}. The gate could be constructed from sequencially applied Toffoli and CNOT gates as shown in \subfigref{fig:peres_gate_decomposed}. The Peres gate inherits the $T$-count and depth of the Toffoli gate since CNOT gate has $T$-count and depth of $0$.

\subsection{Addition and Subtraction Circuits}
\begin{figure}[h]
    \Description[4-bit quantum adder]{Quantum circuit implementation of a 4-bit reversible adder showing the sequence of quantum gates required to perform binary addition without ancilla qubits.}
    \includegraphics{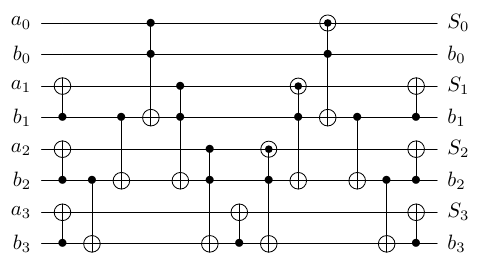}
    \caption{Example 4-bit addition circuit.}
    \label{fig:4_bit_adder}
\end{figure}

The quantum square root algorithm described in the next section also requires T-count optimized addition and subtraction circuits.
The addition circuit used in the current implementation follows the idea from \cite{Thapliyal:2013}. Unlike the original work, this version of the adder omits the overflow qubit. Example of the 4-bit adder is shown in \cref{fig:4_bit_adder}.

The implemented reversible ripple carry adder with no ancilla input qubit produces no garbage and places the result of the calculation in the first register. The algorithm follows six steps described below for two $n$-qubits numbers $a$ and $b$.

\begin{enumerate}
    \item For i=1 to n-1:\\
    Apply the CNOT gate to the qubits $b_i$ and $a_i$ where $a_i$ is the target qubit.
    \item For i=n-1 to 1:\\
    Apply the CNOT gate to the qubits $b_i$ and $b_{i-1}$ where $b_i$ is the target qubit.
    \item For i=0 to n-2:\\
    Apply the Toffoli gate to the qubits $a_i$, $b_i$ and $b_{i+1}$ where $b_{i+1}$ is the target qubit.
    \item For i=n-1 to 0:\\
    If $i=n-1$ apply the CNOT gate to the qubits $b_i$ and $a_i$ where $a_i$ is the target qubit (in the original algorithm the Peres gate is applied, but since the overflow qubit is omitted, the CNOT gate is used instead). Otherwise, apply the Peres gate to the qubits $b_i$, $a_i$ and $b_{i+1}$ such that $b_i$, $a_i$ and $b_{i+1}$ are passed to the inputs $a, b, c$ of the Peres gate respectively.
    \item For i=1 to n-2:\\
    Apply the CNOT gate to the qubits $b_i$ and $b_{i+1}$ where $b_{i+1}$ is the target qubit.
    \item For i=1 to n-1:\\
    Apply the CNOT gate to the qubits $a_i$ and $b_i$ where $b_i$ is the target qubit.
\end{enumerate}

\begin{figure}[h]
    \Description[4-bit quantum subtractor]{Quantum circuit for 4-bit subtraction implemented using NOT gates and an adder circuit, demonstrating how subtraction is performed through addition of inverted values.}
    \includegraphics{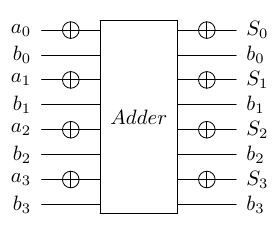}
    \caption{Example of 4-bit subtraction circuit.}
    \label{fig:4_bit_subtractor}
\end{figure}

The subtraction circuit utilizes the property that $a - b = \overline{\overline{a} + b}$. Using this property, subtractor can be designed by inverting the first register, applying the adder and then inverting the first register again \cite{Thapliyal:2016}. The example of such circuit for 4 bit registers is shown in \cref{fig:4_bit_subtractor}.

The $T$ gates are only used in the third and fourth steps of the algorithm and in each step there are $n-1$ Toffoli gates being applied. The total $T$-count then is $(n-1) \cdot 7 + (n-1) \cdot 7$ that can be reduced to $14n - 14$. Since the subtractor doesn't add any additional $T$ gates, the total $T$-count of the subtractor is $14n - 14$ as well.

\subsection{Controlled Addition Circuit}

\begin{figure}[h]
    \Description[4-bit controlled adder]{Quantum circuit for a 4-bit controlled addition operation showing how addition is performed conditionally based on a control qubit state.}
    \includegraphics{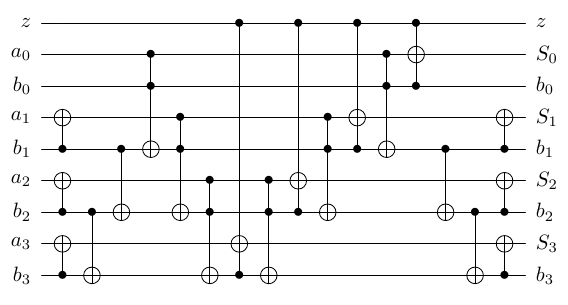}
    \caption{Example 4-bit controlled addition circuit.}
    \label{fig:4_bit_ctrl_adder}
\end{figure}

Another arithmetical operation required by the algorithm is T-Count optimized controlled addition. The implemented version of the controlled addition circuit is theoretically described in \cite{Thaipal:2017}. Unlike the original work, this version of the controlled adder omits the overflow qubits. Example of the 4-bit controlled adder is shown in \cref{fig:4_bit_ctrl_adder}.

The implemented reversible controlled adder with no ancilla input qubit produces no garbage and places the result of the calculation in the first register. The algorithm follows seven steps described below for two $n$-qubit numbers $a$ and $b$ and control qubit $z$.

\begin{enumerate}
    \item For i=1 to n-1:\\
    Apply the CNOT gate to the qubits $b_i$ and $a_i$ where $a_i$ is the target qubit.
    \item For i=n-2 to 1:\\
    Apply the CNOT gate to the qubits $b_i$ and $b_{i+1}$ where $b_{i+1}$ is the target qubit.
    \item For i=0 to n-2:\\
    Apply the Toffoli gate to the qubits $a_i$, $b_i$ and $b_{i+1}$ where $b_{i+1}$ is the target qubit.
    \item Apply the Toffoli gate to the qubits $z$, $b_{n-1}$ and $a_{n-1}$ where $b_{n-1}$ is the target qubit. 
    \item For i=n-2 to 0:\\
    First apply the Toffoli gate to the qubits $a_i$, $b_i$ and $b_{i+1}$ where $b_{i+1}$ is the target qubit.\\
    Then apply the Toffoli gate to the qubits $z$, $b_i$ and $a_i$ where $a_i$ is the target qubit. 
    \item For i=1 to n-2:\\
    Apply the CNOT gate to the qubits $b_i$ and $b_{i+1}$ where $b_{i+1}$ is the target qubit.
    \item For i=1 to n-1:\\
    Apply the CNOT gate to the qubits $b_i$ and $a_i$ where $a_i$ is the target qubit.
\end{enumerate}

The $T$ gates are used in the third, fourth, and fifth steps of the algorithms inside Toffoli gates. In the third step total amount of the Toffoli gates is $n-1$, in the next step there is one Toffoli gate used. In the fifth step there are $2 (n-1)$. Total amount of Toffoli gates is $3n-2$ therefore the T-Count of the circuit is $21n-14$ since the T-Count of the Toffoli gate is $7$.

\section{Quantum Square Root Algorithm}
\label{quantum_square_root}

The quantum circuit presented in \cite{Thapliyal:2018} calculates the integer square root of a number as well as the remainder utlizing the classical non-restoring square root algorithm \cite{Samavi:2008}. The proposed circuit is garbageless, also requires fewer qubits and has a lower $T$-count compared to the existing designs. Consider positive binary value $a$ that has an even bit length $n$. Before computations three registers are initialized: $n$-qubit register $\ket{R}$ that contains $a$, $n$-qubit register $\ket{F}$ set to $1$ and ancillae qubit $\ket{z}$ initialized to $0$. After the computation register $\ket{R}$ will hold the value of the remainder and the $\ket{F}$ register will contain the integer square root on the $a$ on the locations $\ket{F_{n/2-1}}$ through $\ket{F_2}$.

The quantum algorithm is divided into three parts: 1) initial subtraction, 2) conditional addition/subtraction and 3) remainder restoration. 

\subsection{Part 1: Initial Subtraction}
This part occurs once and contains 6 steps.
\begin{enumerate}
    \item Apply the NOT gate on the qubit $R_{n-2}$.
    \item Apply the CNOT gate on the qubit $R_{n-2}$ and $R_{n-1}$ such that $R_{n-1}$ is the target qubit.
    \item Apply the CNOT gate on the qubit $R_{n-1}$ and $F_1$ such that $R_{n-1}$ is the target qubit.
    \item Apply the inverted CNOT gate on the qubit $R_{n-1}$ and the ancilla qubit $z$ such that $R_{n-1}$ is the target qubit.
    \item Apply the inverted CNOT gate on the qubit $R_{n-1}$ and $F_2$ such that $R_{n-1}$ is the target qubit.
    \item Apply the conditioned ADD/SUB such that qubits $R_{n-4}$ to $R_{n-1}$ make the first argument of the ADD/SUB circuit and the qubits $F_0$ to $F_3$ make the second argument while the ancilla qubit $z$ controls the operation performed.
\end{enumerate}

\subsection{Part 2: Conditional Addition or Subtraction}
This part occurs $n/2-2$ times for $i$ from $2$ to $n/2-1$ and is made of 7 steps.
\begin{enumerate}
    \item Apply the inverted CNOT gate on the qubit $z$ and the ancilla qubit $F_1$ such that $F_1$ is the target qubit.
    \item Apply the CNOT gate on the qubit $F_2$ and $z$ such that $z$ is the target qubit.
    \item Apply the CNOT gate on the qubit $R_{n-1}$ and $F_1$ such that $F_1$ is the target qubit.
    \item Apply the inverted CNOT gate on the qubit $R_{n-1}$ and the ancilla qubit $z$ such that $z$ is the target qubit.
    \item Apply the inverted CNOT gate on the qubit $R_{n-1}$ and $F_{i+1}$ such that $F_{i+1}$ is the target qubit.
    \item For $j=i+1$ to $3$\\
    Apply the SWAP gate on the qubits $F_j$ and $F_{j-1}$.
    \item Apply the conditioned ADD/SUB such that qubits $R_{n-1}$ to $R_{n-2 \cdot i-2}$ make the first argument of the ADD/SUB circuit and the qubits $F_{2\cdot i+1}$ to $F_0$ make the second argument while the ancilla qubit $z$ controls the operation performed.
\end{enumerate}

\subsection{Part 3: Remainder Restoration}
The last part occurs only once and contains 9 steps.
\begin{enumerate}
    \item Apply the inverted CNOT gate on the qubit $z$ and $F_1$ such that $F_1$ is the target qubit.
    \item Apply the CNOT gate on the qubit $F_2$ and $z$ such that $z$ is the target qubit.
    \item Apply the inverted CNOT gate on the qubit $R_{n-1}$ and $z$ such that $z$ is the target qubit.
    \item Apply the inverted CNOT gate on the qubit $R_{n-1}$ and $F_{n/2+1}$ such that $F_{n/2+1}$ is the target qubit.
    \item Apply the NOT gate on the qubit $z$.
    \item Apply the controlled addition on the registers $R$, $F$ and $z$ such that if ancilla qubit $z$ has value $1$ the $R$ register will hold the value $R+F$ and $F$ will be unchanged. If $z$ is $0$ both registers will be unchanged.
    \item Apply the NOT gate on the qubit $z$.
    \item For $j=n/2 + 1$ to $3$\\
    Apply the SWAP gate on the qubits $F_j$ and $F_{j-1}$.
    \item Apply the CNOT gate on the qubit $F_2$ and $z$ such that $z$ is the target qubit.
\end{enumerate}

After the last step the qubits $\ket{F_{n/2+1}}$ through $\ket{F_2}$ will contain the integer square root of $a$. And the register $\ket{R}$ will hold the remainder.

\section{Implementation in Qrisp}
\label{implementation}

Implementation of the non-restoring integer square root algorithm is done as a reversible quantum circuit using the Qrisp framework. Each stage of the classical algorithm corresponds to a specific quantum sub-circuit, composed and executed on quantum registers. Each component of the implementation is detailed below. 

\subsection{Peres Gate}
To facilitate arithmetic operations, a three-qubit \emph{Peres gate} is defined (the introduced gate that combines Toffoli and CNOT gates). The Qrisp code in \texttt{peres\_gate} function constructs this gate by applying a CCX (Toffoli) followed by a CX (CNOT), and converts the circuit into a reusable gate. 

\begin{lstlisting}[language=Python]
def peres_gate() -> Operation:
    qc = QuantumCircuit(3)
    qc.ccx(0, 1, 2)
    qc.cx(0, 1)
    return qc.to_gate(name="PERES")
\end{lstlisting}

\subsection{Reversible Addition Circuit}
Using the Peres gate, an $n$-bit ripple-carry adder circuit is implemented in the function named \texttt{add\_circuit(n)}, which returns a reversible gate "ADD" acting on $2n$ qubits. The gate adds two $n$-qubit numbers while leaving the second register unchanged and storing the sum $A+B$ in the first register. 

The code below follows a six-step operation. Registers $A$ and $B$ are defined as lists of qubit positions, and comments "Step1"–"Step6" correspond to the stages of the addition algorithm described in the \nameref{basics} section.
\begin{lstlisting}[language=Python]
def add_circuit(n: int) -> Operation:
    # <... Preparing the circuit ...>
    # Step 1
    for i in range(1, n):
        qc.cx(B[i], A[i])
    # Step 2
    for i in range(n-2, 0, -1):
        qc.cx(B[i], B[i+1])
    # Step 3
    for i in range(0, n-1):
        qc.ccx(B[i], A[i], B[i+1])
    # Step 4
    for i in range(n-1, -1, -1):
        if i == n - 1:
            qc.cx(B[i], A[i])
        else:
            qc.append(
                peres_gate(), 
                [B[i], A[i], B[i+1]])
    # Step 5
    for i in range(1, n-1):
        qc.cx(B[i], B[i+1])
    # Step 6
    for i in range(1, n):
        qc.cx(B[i], A[i])
    return qc.to_gate(name="ADD")
\end{lstlisting}

\lstinline[language=Python]{# <... Preparing the circuit ...>} is used to avoid boilerplate code and indicates the part of the code that creates a circuit and defines registers.

\subsection{Controlled Addition/Subtraction}
Some steps in the non-restoring square root algorithm requires usage of the addition and subtraction opearations based on a condition (the sign of the current remainder). This step is implemented in the \texttt{ctrl\_add\_sub\_circuit(n)} function that constructs a reversible $(2n+1)$-qubit gate.

\begin{lstlisting}[language=Python]
def ctrl_add_sub_circuit(n: int) -> Operation:
    # <... Preparing the circuit ...>
    for i in A:
        qc.cx(z, i)
    qc.append(add_circuit(n), A + B)
    for i in A:
        qc.cx(z, i)
    return qc.to_gate(name="CTRL ADD/SUB")
\end{lstlisting}
As shown in the code, the ADD circuit is always applied and the only thing that changes based on the condition is the sign of the first argument $A$.

\subsection{Controlled Addition}

The last step of the original algorithm utilizes the controlled addition circuit to make final adjustments of the arguments based on the remainder's sign. The implementation of the circuit is located in \texttt{ctrl\_add\_circuit} function. If the controlled qubit is $1$, the function calculates the sum of the two numbers and places the result in the first register $A$ (the second register remains unchanged); otherwise, both registers stay unchanged. The code below follows the seven steps of the described controlled adder.

\begin{lstlisting}[language=Python]
def ctrl_add_circuit(n: int) -> Operation:
    # <... Preparing the circuit ...>
    # Step 1
    for i in range(1, n):
        qc.cx(B[i], A[i])

    # Step 2
    for i in range(n - 2, 0, -1):
        qc.cx(B[i], B[i+1])

    # Step 3
    for i in range(0, n - 1):
        qc.ccx(A[i], B[i], B[i+1])

    # Step 4
    qc.ccx(z, B[n-1], A[n-1])

    # Step 5
    for i in range(n-2, -1, -1):
        qc.ccx(A[i], B[i], B[i+1])
        qc.ccx(z, B[i], A[i])

    # Step 6
    for i in range(1, n-1):
        qc.cx(B[i], B[i+1])

    # Step 7
    for i in range(1, n):
        qc.cx(B[i], A[i])
    return qc.to_gate(name="CTRL ADD")
\end{lstlisting}

\subsection{Initial Subtraction Stage}
The algorithm begins by an initial subtraction on the most significant bits to establish the first partial remainder. The function \texttt{part1\_circuit(n)} returns a $(2n+1)$-qubit operation "PART 1" that prepares the remainder register $R$ and the result register $F$ for the iterative process. The code below shows this initialization step.

\begin{lstlisting}[language=Python]
def part1_circuit(n: int) -> Operation:
    # <... Preparing the circuit ...>
    # Step 1
    qc.x(R[n-2])
    # Step 2
    qc.cx(R[n-2], R[n-1])
    # Step 3
    qc.cx(R[n-1], F[1])
    # Step 4
    qc.append(zcx(), [R[n-1], z])
    # Step 5
    qc.append(zcx(), [R[n-1], F[2]])
    # Step 6
    qc.append(
        ctrl_add_sub_circuit(4),
        [z, R[n-4], R[n-3], 
        R[n-2], R[n-1], F[0], 
        F[1], F[2], F[3]])
    return qc.to_gate(name="PART 1")
\end{lstlisting} 
First three steps are implemented using Qrisp functions \texttt{x(qubit)} and \texttt{cx(ctrl, target)} that implement the NOT and CNOT gates respectively. The fourth and fifth steps require inverted CNOT operation (zero-controlled CNOT gate) that is implemented by \texttt{zcx()} function. The function works as a wrapper for the \texttt{XGate().control(ctrl\_state=0)} gate. The gate performs CNOT operation on the target qubit if the control qubit is in state $\ket{0}$. The last step uses the \texttt{ctrl\_add\_sub\_circuit(4)} function that implements the controlled addition/subtraction on the qubits specified in the argument (qubits $z$, $R[n-4]$ to $R[n-1]$ and $F[0]$ to $F[3]$).

\subsection{Conditinal Addition/Subtraction Stage}
After initialization, the algorithm processes the remaining bits of the input in pairs. The function \texttt{part2\_circuit(n)} generates the looped circuit that handles each subsequent pair of bits, using the control logic to decide on addition or subtraction at each step. The pseudocode is essentially a loop that, for each iteration $i$, prepares the control signals and then applies a controlled add/subtract on an expanding portion of the registers. The implementation is shown below. 
\begin{lstlisting}[language=Python]
def part2_circuit(n: int) -> Operation:
    # <... Preparing the circuit ...>
    for i in range(2, n // 2):
        # Step 1
        qc.append(zcx(), [z, F[1]])
        # Step 2
        qc.cx(F[2], z)
        # Step 3
        qc.cx(R[n-1], F[1])
        # Step 4
        qc.append(zcx(), [R[n-1], z])
        # Step 5
        qc.append(zcx(), [R[n-1], F[i+1]])
        # Step 6
        for j in range(i + 1, 2, -1):
            qc.swap(F[j], F[j-1])
        # Step 7
        R_sum_qubits = [R[j] for j in
                 range(n - 2 * i - 2, n)]
        F_sum_qubits = [F[j] for j in 
                 range(0, 2 * i + 2)]
        l = len(R_sum_qubits)
        qc.append(
            ctrl_add_sub_circuit(l),
            [z] + R_sum_qubits + F_sum_qubits)
    
    return qc.to_gate(name="PART 2")
\end{lstlisting}
First five steps are implemented in the similar manner as in the previous part. The sixth step uses the \texttt{swap(q1, q2)} function to swap the argument qubits. In the last step the \texttt{ctrl\_add\_sub\_circuit()} function is appended to the circuit with the control qubit $z$.

\subsection{Remainder Restoration Stage}
After processing all pairs of bits, the algorithm may end in a state where the last operation was a subtraction, potentially leaving a negative remainder. The final step is to restore a correct non-negative remainder. The function \texttt{part3\_circuit(n)} produces a sub-circuit "PART 3" that conditionally adds back the last subtracted value. Its implementation is given below. 
\begin{lstlisting}[language=Python]
def part3_circuit(n: int) -> Operation:
    # <... Preparing the circuit ...>
    # Step 1
    qc.append(zcx(), [z, F[1]])
    # Step 2
    qc.cx(F[2], z)
    # Step 3
    qc.append(zcx(), [R[n-1], z])
    # Step 4
    qc.append(zcx(), [R[n-1], F[n//2+1]])
    # Step 5
    qc.x(z)
    # Step 6
    qc.append(ctrl_add_circuit(n), 
        [z] + R[:] + F[:])
    # Step 7
    qc.x(z)
    # Step 8
    for j in range(n//2 + 1, 2, -1):
        qc.swap(F[j], F[j-1])
    # Step 9
    qc.cx(F[2], z)

    return qc.to_gate(name="PART 3")
\end{lstlisting}
The operations used in the last part of the algorithm are implemented in a similar way as previous parts.

\subsection{Assembling the Square Root Circuit}
The top-level function \texttt{square\_root\_circuit(n)} composes the full quantum circuit for the integer square root by concatenating the three stages described above. As shown below, it simply appends the gates for initial subtraction, the iterative conditional add/subtract, and remainder restoration in sequence on a common set of registers $R$, $F$, and $z$. 
\begin{lstlisting}[language=Python]
def square_root_circuit(n: int) -> Operation:
    # <... Preparing the circuit ...>
    part1 = part1_circuit(n)
    part2 = part2_circuit(n)
    part3 = part3_circuit(n)
    qc.append(part1, R[:] + F[:] + [z])
    qc.append(part2, R[:] + F[:] + [z])
    qc.append(part3, R[:] + F[:] + [z])
    return qc.to_gate(name="ISQRT")
\end{lstlisting} 
This assembled gate (named "ISQRT") acts on $2n+1$ qubits, where $n$ is chosen based on the input size. In our implementation, $n$ is determined as the smallest even number of qubits sufficient to represent the input number $a$ in binary (with two extra bits if needed to accommodate the algorithm's grouping of bits). The ISQRT circuit can then be applied to quantum registers representing the input and will produce the integer square root in the $F$ register and the remainder in the $R$ register upon measurement. 

\subsection{Executing the Circuit in Qrisp}
Finally, to use this circuit within Qrisp, we allocate quantum registers and run a quantum session. Qrisp provides the class \texttt{QuantumFloat(bit\_length, exponent)} to represent a $n$-qubit quantum number and \texttt{QuantumSession()} to simulate circuit execution. The \texttt{isqrt} function as input gets a 2's complement quantum number $R$ with an even number of qubits and as a result returns square root the input and transforms $R$ register into the reminder. 

We prepare two quantum registers: $F$ (result), and $z$ (control flag). Initially, $F$ is set to 1 (as required by the algorithm's initial conditions), and $z$ to 0. We then append the \texttt{ISQRT} gate to a session and execute it. After execution, the square root is in qubits $F_{n/2 + 1}$ to $F_2$, so we also need to shift $F$ by 2, to correctly return the square root.

\begin{lstlisting}[language=Python]
def isqrt(R: QuantumFloat) -> QuantumFloat:
    n = R.size
    F = QuantumFloat(R.size, 0, name="F")
    z = QuantumFloat(1, 0, name="z")
    F[:] = 1
    z[:] = 0

    qs = QuantumSession()
    qs.append(square_root_circuit(n),
        R[:] + F[:] + z[:])
    
    qs.x(F[0])
    for i in range(2, n // 2 + 2):
        qs.swap(F[i], F[i - 2])

    return F
\end{lstlisting} 

As a result of this function, we obtain a quantum number whose value is the square root. Meanwhile, the input parameter $R$,
which originally held the square, now carries the remainder of the computation.

In summary, the described implementation in Qrisp provides a complete quantum circuit for the integer square root using the non-restoring method. Each logical step of the classical algorithm is mirrored by a reversible quantum operation, and the Qrisp framework allows us to combine these into a single coherent quantum circuit (\texttt{ISQRT}) that can be applied and tested on arbitrary input values. The implementation illustrates how high-level quantum programming constructs (like controlled operations and modular circuit composition) can be used to realize complex arithmetic algorithms on quantum hardware.

\section{Validation}
\label{validation}

To test the implemented circuit, we run it for a range of different input values. The simple test code is shown below.
\begin{lstlisting}[language=Python]
qa = QuantumFloat(n, 0, signed=True)
qa[:] = a
qf = isqrt(qa)
# getting a state of the variable that has probablity 1
root = get_value(qf)
remainder = get_value(qa)
print(f'a = {a}, root = {root}, remainder = {remainder}')
\end{lstlisting}

Functional verification confirmed the correct operation of the circuit across all test cases -- output of the tests for numbers from $80$ to $87$ is shown below.
\begin{lstlisting}
$ python tests.py
a = 6, root = 2, remainder = 2
a = 7, root = 2, remainder = 3
a = 8, root = 2, remainder = 4
a = 9, root = 3, remainder = 0
a = 10, root = 3, remainder = 1
a = 11, root = 3, remainder = 2
a = 12, root = 3, remainder = 3
a = 13, root = 3, remainder = 4
a = 14, root = 3, remainder = 5
a = 15, root = 3, remainder = 6
a = 16, root = 4, remainder = 0
\end{lstlisting}

It is worth mentioning that the authors of the quantum algorithm estimated the T-count of their proposed square root circuit as \( \frac{7}{2}n^2 + 21n - 28 \)~\cite{Thapliyal:2018}. In our implementation, the T-count remains unchanged, as the only components of the algorithm that utilize T gates are \textsc{CTRL ADD} and \texttt{CTRL ADD/SUB} circuits. Our version of the circuits achieves the same T-count as the one analyzed in the original work. 

In Qrisp it is possible to validate count of every operation used in the circuit by calling \texttt{count\_ops} function on the analyzed circuit. In our implementation, every part of the algorithm is divided into different circuit, therefore to get count of used operations the code should be modified in a way that every part of the algorithm is applied to the same \texttt{QuantumCircuit}. It is also possible to view total amount of qubits used by a circuit by calling \texttt{num\_qubits} function on the circuit. Table \ref{tab:validation} shows the computed amount of qubits and the T-Count used by the square root circuit for different input sizes, as well as their theoretical estimation shown by authors in \cite{Thapliyal:2018}.

\begin{table}[h!]
\centering
\caption{Comparing computed total amount of qubits (Total qubits (C)) and the T-Count (T-Count (C)) with their theoretical estimations (Total qubits (E) and T-Count (E) respectively) for different input sizes (n).}
\begin{tabular}{|c|c|c|c|c|}
\hline
\textbf{n} & \textbf{Total qubits (C)} & \textbf{Total qubits (E)} & \textbf{T-Count (C)} & \textbf{T-Count (E)} \\
\hline
6 & 13 & 13 & 224 & 224 \\
8 & 17 & 17 & 364 & 364 \\
10 & 21 & 21 & 532 & 532 \\
12 & 25 & 25 & 728 & 728 \\
14 & 29 & 29 & 952 & 952 \\
16 & 33 & 33 & 1204 & 1204 \\
\hline
\end{tabular}
\label{tab:validation}
\end{table}

\section{Conclusion}
\label{conclusion}
In this work, we have successfully implemented the T-count optimized non-restoring quantum square root algorithm proposed by Thapliyal \cite{Thapliyal:2018} using the Qrisp quantum programming framework. This represents the first complete implementation of this algorithm in a high-level quantum development environment, demonstrating its practical realizability beyond theoretical analysis.

Our implementation validates the theoretical resource estimates from the original work, achieving a T-count of $14n-14$ and T-depth of $5n+3$ for an $n$-bit input. The modular design approach enabled by Qrisp allowed us to construct the circuit from reusable components including reversible adders, subtractors, and conditional logic blocks, each built from fundamental quantum gates and the Peres gate primitive.

A key contribution of this work is demonstrating how quantum programming frameworks like Qrisp can facilitate the implementation of arithmetic algorithms. The framework's support for high-level constructs such as controlled operations, modular circuit composition, and quantum register management significantly simplified the development process compared to gate-level circuit construction.
This high-level approach represents a significant step toward the practical implementation of algorithms on large-scale quantum computers.

\section{Materials \& Code Availability}
The implementation code is available in the GitHub repository \cite{code}.

\section{Acknowledgments}
This work was supported by the EU Horizon Europe Framework Program under Grant Agreement no. 101119547 (PQ-REACT).

\bibliographystyle{ACM-Reference-Format}
\bibliography{base}

\end{document}